\documentclass{moriond}

\usepackage{amsmath,graphicx,epsfig}
\usepackage{amssymb}

\def\be{\begin{equation}}
\def\ee{\end{equation}}
\def\bea{\begin{eqnarray}}
\def\eea{\end{eqnarray}}

\def\prn#1{{\left(#1\right)}}
\def\Msun{{\ensuremath{M_\odot}}}
\def\mr{\mathrm}

\newcommand{\Photo}{\includegraphics[height=35mm]{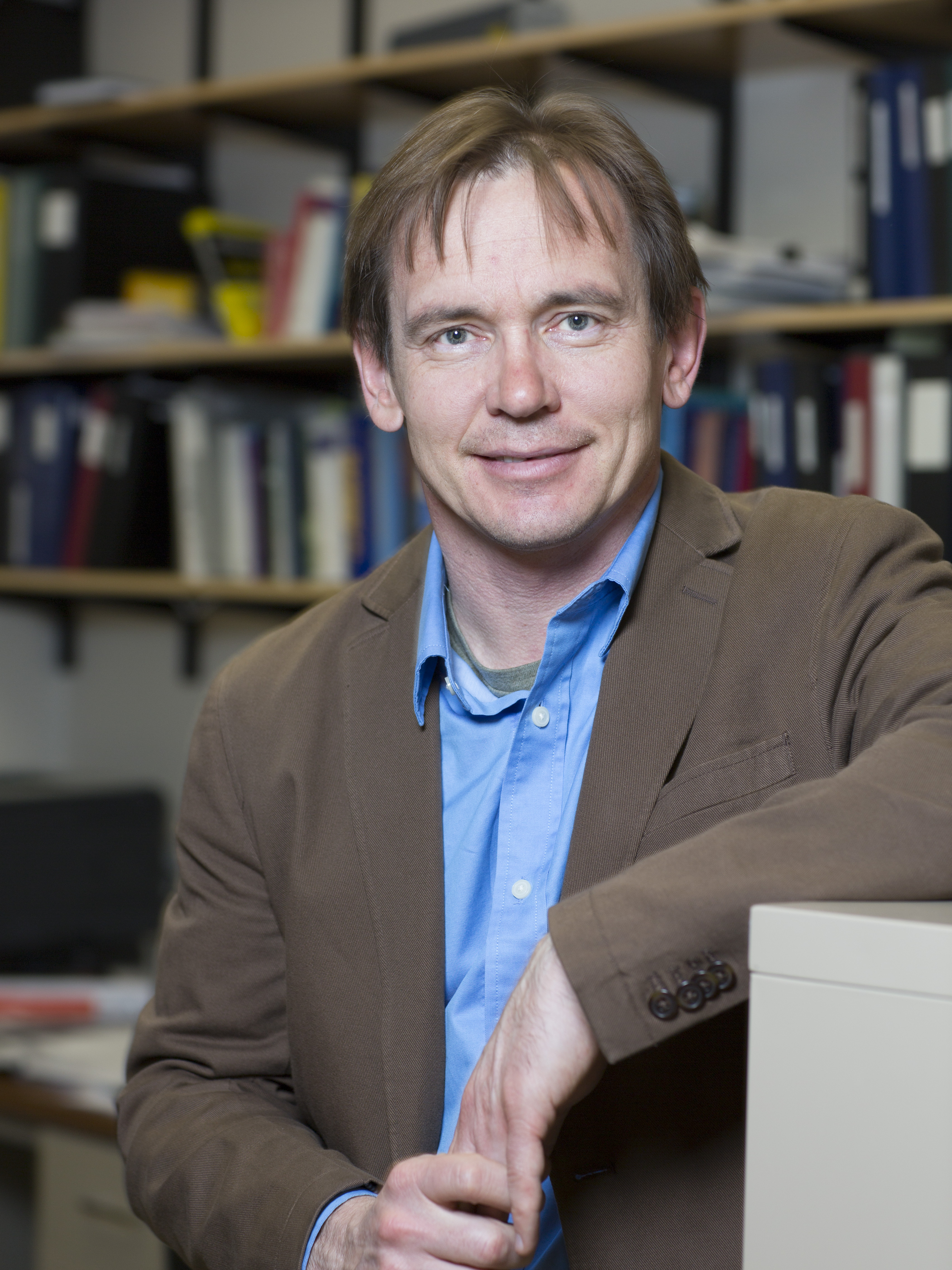}}

\begin{document}
\vspace*{4cm}

\title{Quantum gravity unchained: Atomic sensors as exotic field telescopes in multi-messenger astronomy}

\author{ A. Derevianko }

\address{Department of Physics, University of Nevada, Reno, Nevada 89557, USA}

\maketitle
\abstracts{
We propose a novel, exotic physics, modality in multi-messenger astronomy.   We are interested in a {\em direct} detection of exotic fields emitted by the mergers. This approach must be contrasted with the {\em indirect} detection strategies, e.g., based on minute exotic-physics induced changes in gravitational wave spectral features. While our strategy seems to be overly optimistic, the numbers do work out. The numbers work out because of (i) the exquisite sensitivity of atomic quantum sensors and because of (ii) the enormous amounts of energy released in the mergers. Bursts of exotic fields may, for example, be produced 
during the coalescence of  black hole  singularities, releasing quantum gravity messengers  per the title of this contribution. To be detectable by the precision atomic sensors, such fields must be ultralight and ultra-relativistic and we refer to them as exotic low-mass fields (ELFs). Since the fields are massive,  the group velocity of ELF bursts is smaller than the speed of light. Thereby the ELF bursts lag behind the gravitational waves. Then  LIGO or other gravitational wave observatories would provide a trigger for networks of precision atomic sensors that can listen for the feeble ELF signals. We characterize ELF signatures in the sensors. ELFs  would imprint a characteristic  anti-chirp signal across the sensor network. This contribution to Moriond-Gravity proceedings summarizes salient points of our previous publication [Dailey et al., Nature Astronomy 5, 150 (2021)]. I aim at a discussion that is informal and accessible yet grounded in quantitative estimates.
}

\section{Introduction}

Since the initial discovery of gravitational waves (GW) by LIGO in 2015~\cite{LIGOfirstObservation2016}, there were multiple observations of GW arrivals at the Earth. Most of these GW result from mergers of a pair of black holes  However, on  August 17, 2017, a new class of GW  events was discovered: the binary neutron star merger GW170817~\cite{LIGOVirgo-NeutronStarMerger2017}.  That was the first astrophysical source detected in both gravitational waves and multiwavelength electromagnetic radiation. This  event has generated a considerable excitement in the astrophysics community, ushering the era of multi-messenger astronomy~\cite{multimessenger2017}.  A $\sim 100\, \mathrm{s}$ long detected GW signal  was followed by a short gamma-ray burst.  An optical transient was subsequently found in the host galaxy NGC 4993 at a distance of 40 megaparsecs ($\sim$ 100 million light-years). The source was detected in  a comprehensive observational campaign~\cite{Coulter2017,Hallinan2017,Soares-Santos2017,Troja2017,Margutti2018,Mooley2018,Ruan2018} across the electromagnetic spectrum -- in the X-ray, ultraviolet, optical, infrared, and radio bands over hours, days, and weeks. Early detections were obtained within one day of the GW trigger~\cite{Coulter2017,Soares-Santos2017}. 

Our paper~\cite{dailey2020ELF.Concept} extends the gravitational and electromagnetic modalities of multi-messenger astronomy to exotic (beyond the Standard Model of elementary particles) fields. We are interested in a {\em direct} detection of exotic fields emitted by the mergers. This approach must be contrasted with {\em indirect} detection strategies, e.g., based on minute exotic-physics induced changes in GW spectral features. While our strategy seems to be overly optimistic, the numbers do work out.
The numbers work out because of (i) the exquisite sensitivity of atomic sensors and because of (ii) the enormous amounts of energy released in the mergers.

\begin{figure}
\center
\includegraphics[width=0.75\textwidth]{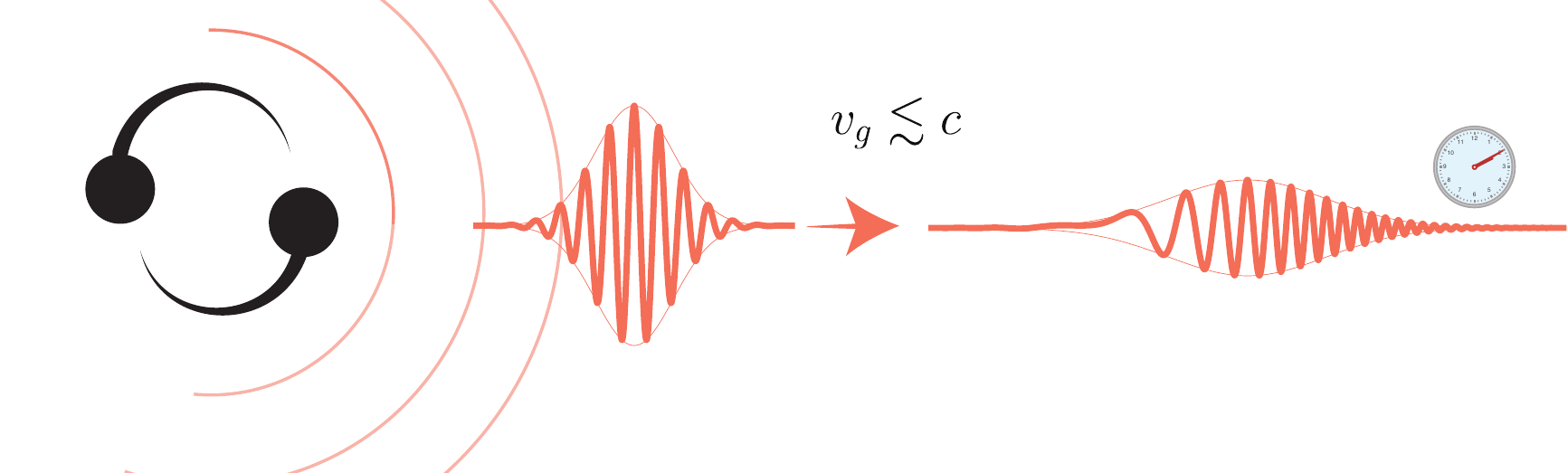}
\caption[]{ 
An emitted ELF burst propagates with  the group velocity $v_g \lesssim c$ to the detector, lagging behind the GW burst.  
Since the more energetic ELF components propagate faster, the arriving ELF wavepacket exhibits  a characteristic  frequency anti-chirp. From our paper\cite{dailey2020ELF.Concept}.
}
\label{Fig:Cartoon}
\end{figure}

Fig.~\ref{Fig:Cartoon} summarizes our idea.  A merger  emits both the GW and an exotic  low-mass fields (ELFs) bursts. Since the ELF is massive, the burst propagates at a group velocity smaller than the speed of light. Thereby, the ELF pulse lags behind the GW burst. Because of the dispersion, the ELF burst tends to spread out as it travels. More energetic components of the  ELF wave-packet travel faster, imprinting a universal anti-chirp signature in the quantum sensor data (in this case an atomic clock).

As to the exotic, beyond the standard model, physics modality, we focus on ultralight (yet non-zero mass) bosonic fields as the messenger. 
Indeed, ultralight bosons (masses $\ll {\rm 1~eV}$) are ubiquitous in various new physics scenarios. Such exotic fields are posited, for example, in explaining the nature of dark matter and dark energy, the hierarchy problem, the strong CP problem, and in  quantum theories of gravity. We presupose that  bursts of such ELFs could be generated by cataclysmic astrophysical events such as black hole or neutron star mergers\cite{bini2017deviation,baumann2019probing}.

Quantum sensors, such as atomic clocks and magnetometers, are sensitive to gentle perturbations of internal degrees of freedom (energy levels, spins, etc.) by coherent, classical waves. Then to be detectable by the atomic sensors, the astrophysical source must produce coherent ELF waves with high mode occupation number. The mode occupation number can be diluted as the spherical wave-front expands as it travels and the condition must be satisfied all the way to the sensors.~\footnote{Emitted ELFs are indeed copious ($ \gtrsim 10^{70}$ for $\Delta E \sim 0.1 M_\odot c^2$ and $\omega_0 = 2\pi \times 10 \,\mathrm{kHz}$).}

There is a wide variety of speculative scenarios for coherent ELF production, enumerated in our original publication\cite{dailey2020ELF.Concept};
these range from the scalar-tensor gravity to stripping away boson clouds coherently built up around black holes (BH).
My personal favorite  is due to quantum gravity of BH singularities. Much of the underlying physics of coalescing singularities in BH mergers remains unexplored as it requires understanding of the as yet unknown theory of quantum gravity\cite{Loeb2018-BH-signularities}. Then the ELF burst would emerge from the merger as a  quantum gravity messenger. Hence the title of my contribution ``Quantum gravity unchained''.

Considering that much of the underlying physics remains to be discovered, we take a pragmatic  approach  based on energy arguments. We assume, based on the  uncertainties for the LIGO events,
 that $\Delta E \sim M_\odot c^2$ of energy  can be released in the form of ELFs from a black hole merger and $\Delta E \sim 0.1 \,M_\odot c^2$ in a neutron star merger.
 In other words, enormous (compared to terrestrial experiments) amounts of energy can be released into the ELF channel. 
 
\section{Propagation of the ELF burst and its signature at the sensor}
 
As an illustration, let's assume that the emitted ELF is a spinless field. Klein-Gordon equation governs quantum physics of such bosons; it admits the usual wave solutions. An emitted ELF  propagates out as a spherical wave with the conventional relativistic energy-momentum dispersion relation
 \begin{equation}
\omega(k)=\sqrt{(c k)^2+\Omega_c^2},
\label{Eq:dispersion}
\end{equation}
where the  Compton frequency $\Omega_c = mc^2 /\hbar$ depends on  the ELF mass $m$.  The energy and momentum of  ELF quanta are given by $\varepsilon = \hbar \omega$ and $p=\hbar k$. A linear combination of outgoing spherical waves of different frequencies forms an outward propagating wavepacket.
A duration of the burst $\tau_0$ implies a frequency (or energy) spread $\Delta \omega= 1/ \tau_0$. Given the fixed spread and central frequency $\omega_0$ of the ELFs,  the Bayesian minimum entropy principle implies that the wavepacket has a Gaussian shape. The amplitude of the wavepacket is determined by the total energy $\Delta E \sim M_\odot c^2$  in the ELF channel.
This will be our generic assumption, which, of course, may be fine-tuned for a specific production mechanism. Even in this case, our observations about the gross features of the expected ELF signature at the sensors hold true. 

 Formally the ELF can be described as a superposition of spherically-symmetric waves:
$\phi_k(r,t) = \frac{A_k}{r}\cos\prn{ k  r - \omega(k) t  + \theta_k},$  
where $r$ is the radial coordinate, $A_k$, $\theta_k$, $k$, and $\omega$ are the ELF amplitudes, phases, wavevectors, and frequencies, respectively. The initial Gaussian wavepacket can be represented as a linear Fourier combination of these spherical waves. 
Individual components $\phi_k(r,t)$ propagate with different phase velocities $k/\omega(k)$. 
As the wavepacket evolves, its envelope will change over time exhibiting dispersion with the center of the envelope moving at the group velocity.

The dispersion relation~(\ref{Eq:dispersion}) has important implications for the propagation  of the ELF  wavepacket. 
In particular, we are interested in 
\begin{enumerate}
\item{ The group velocity $v_g=\partial \omega  /\partial k $ as  it determines the time  delay $\delta t$ of the  ELF signal  with  respect to the  GW burst moving at the speed of light;}
\item{ The duration $\tau$  and  the  amplitude of the ELF wavepacket at the sensor;} 
\item{Time dependence of instantaneous ELF wavepacket frequency at the sensor.}
\end{enumerate}

The dispersion relation~\eqref{Eq:dispersion} translates into an ``internal'' index of refraction
\begin{equation}
n\left(  \omega\right)  =\frac{k}{\omega}=\sqrt{1-\frac{\Omega_c^{2}}{\omega^{2}}} \,, \label{Eq:IndRefraction}
\end{equation}
mapping the problem of the ELF wavepacket propagation into well-understood wave-propagation in electrodynamics~\cite{JacksonEM}.  In particular, the center of the wavepacket moves at the group velocity $v_g\le c$. We need to require that $v_g \sim c$ as we want to correlate the ELF signal with a specific LIGO GW trigger.  
Indeed, LIGO-detected GWs must traverse the vast billion-light year distances and if we want to limit the GW-ELF time delay $\delta t$  to less than a week, ELFs must be ultra-relativistic. Then the ELF  central frequency $\omega_0$ and wavevector   $k_0$ are  related  by photonic dispersion $\omega_0 \approx c k_0$ and the index of refraction in this limit is
\begin{equation}
n\left(  \omega\right) \approx 1- \frac{\Omega_c^{2}}{2\omega^{2}} \,. \label{Eq:IndRefractionSimplified}
\end{equation}

Then we can consider a propagation of a Gaussian pulse with this simplified index of refraction. The  solution (c.f. Ref.\cite{JacksonEM} and detailed derivation in Ref.\cite{dailey2020ELF.Concept}) for a field at the sensor a distance $R$ away from the progenitor reads 
\begin{align}
\phi(R,t) \approx & \frac{1}{R} \prn{\frac{c\Delta E}{2\pi^{3/2} \omega_0^2 \tau}}^{1/2} \exp\prn{-\frac{(t-t_s)^2}{2 \tau^2}} \nonumber\\
                  & \times \cos\prn{ \omega_0 (t-t_s) - \frac{\omega_0}{4\delta t}(t-t_s)^2 }\,,
\label{Eq:DispersionModeldE}
\end{align}
where $t_s = t_\mathrm{GW} +  \delta t $ is the time of arrival of the center of the pulse to the sensor. It is offset by  $t_\mathrm{GW} = R/c$, the time of arrival of the GW messenger. 
In contrast to dispersionless spherical waves, the  field amplitude at the sensor  $\phi(R,t)$ scales  as  $1/R^{3/2}$,  reflecting the  pulse dispersion. The duration  $\tau$ of the ELF pulse at the sensor can be estimated as  $\tau  \sim R  \Delta  v_g/c^2$,  where the spread in group  velocities  $\Delta  v_g/c \approx  \partial^2\omega/\partial k^2/\tau_0$.  This relates the signal duration and GW-ELF time delay
\begin{equation}
 \tau \approx  2  \delta t \,    /(\omega_0 \tau_0) \, .
 \label{Eq:duration}
\end{equation}
Note that the instantaneous ELF frequency is time-dependent, $\omega(t)  = \prn{ 1 - (t-t_s) /(2\delta t)}\omega_0$, exhibiting a  frequency ``anti-chirp" at the sensor.  The waveform, Eq.~\eqref{Eq:DispersionModeldE}, is shown in Fig.~\ref{Fig:Cartoon}. The slope of the anti-chirp is given by $d\omega/dt = - 1/\prn{\tau \tau_0}=-\omega_0/\prn{2\delta t}$. This reflects a qualitative fact that the more energetic (or higher frequency) components have a higher phase velocity $k/\omega$. Thereby the higher-frequency ELF components arrive earlier at the sensor.

It is worth noting that the waveform~\eqref{Eq:DispersionModeldE} is consistent with conservation of energy: A shell of radius $R$ and of width $\tau c$ contains the released energy $\Delta E$ in the ELF channel.

\section{Catching an ELF}
Now we know how an ELF pulse looks like at a sensor. How can we detect them? Have the ELFs, if they do exist, been already constrained as they can participate in other processes?

I will focus on atomic clocks and optical cavities; discussion of atomic magnetometers can be found in Ref.\cite{dailey2020ELF.Concept}.
First of all, there is only a range of frequencies that we are sensitive to. A typical atomic clock sampling rate is below 1Hz, while for cavities it is below 10 kHz. This limits $\omega_0$ from above. 
These  frequencies fix energies $\varepsilon_0 =\hbar \omega_0$  of detectable ELFs  to  below $10^{-14} \,  \mathrm{eV}$ for clocks and $10^{-10} \,  \mathrm{eV}$ for cavities.
Practically, we are limited to a duration of the pulse at the sensor to about a month; this limits $\omega_0$ from below.

Atomic clocks work by locking the frequencies of tunable sources of microwave or optical radiation to frequencies of atomic transitions. The atoms (quantum oscillators) are well protected from the environment or the environmental perturbations are well characterized. Telling time then relies on counting the number of oscillations of the radiation at the source and multiplying it by the period of the oscillation determined by the atomic transition frequency.  If an exotic field pulls on the atomic energy levels, the change in the clock frequency can be determined, for example, by comparing it with another clock that responds to the exotic fields in a different fashion, because it is either remote and not affected by the exotic field or is using a different atomic species. 

The most obvious way how the atomic frequencies can be affected is through the variation of fundamental constants. For example, the optical frequencies are proportional to the Rydberg constant $m_e c^2 \alpha^2$. So if $\alpha$ varies, so do the clock frequencies. In clock comparisons, however, this overall scaling factor cancels out, and we need to consider relativistic effects, that bring in additional $\alpha$ dependence that differs for different atomic species. If the clocks are compared remotely, however, the clocks may use the same species.

Variations of fundamental constants may be induced by phenomenological portals. The least constrained is the quadratic portal with a Lagrangian density
\begin{equation}
\mathcal{L}^{(2)}_\mathrm{clk} = \prn{ -\Gamma_{m_e}  m_e c^2   \bar{\psi_e} \psi_e  + \Gamma_\alpha F_{\mu \nu}^2 /4 } \phi^2.
\end{equation}
Here  $\psi_e$ is the  electron bi-spinor,  $F_{\mu\nu}$ is the Faraday tensor,  and $\Gamma$'s are coupling constants. Quadratic interactions appear naturally for ELFs possessing either $Z_2$ or $U(1)$ intrinsic symmetries. The first term varies  the electron mass and the second -- fine-structure constant. Typically the coupling constants are reframed in terms of energy scales $\Lambda_X \equiv 1/\sqrt{|\Gamma_X|}$.

Detailed analysis~\cite{dailey2020ELF.Concept}, based on the excess power statistic technique, sets the following limits on sensitivity 
\begin{eqnarray}
\Lambda_X & \lesssim & 0.1 \, 
\prn{ \frac{\sqrt{N_s} |K_X| }{\sigma_y(\Delta_t) }}^{1/2}
 \prn{  \frac{c}{R\omega_0}} 
  \prn{\frac{\hbar^2 \Delta E^2} {\Delta_t \tau}}^{1/4}.\label{Eq:LambdaConstraint}
\end{eqnarray}
Here $K_X$ is the sensitivity coefficient to a variation in fundamental constant $X=\{m_e,\alpha, \ldots\}$,  $\Delta_t$ is the sensor  sampling time interval, $\sigma_y(\Delta_t)$ is the dimensionless clock Allan deviation over sampling interval $\Delta_t$  for  fractional frequency excursions, and $N_s$ is the number of sensors.  Effectively, we require that the signal-to-noise ratio exceeds a 95\% confidence level.

Astrophysical observations and laboratory experiments set constraints on the coupling strengths between ELFs and standard model particles and fields~\cite{safronova2018search}. 
Using the above sensitivity estimates, we
 find that the current generation of atomic clocks is sensitive to quadratic  portals $\mathcal{L}^{(2)}$ 
 
 \begin{figure}[ht!]
\center
\includegraphics[width=0.5\columnwidth]{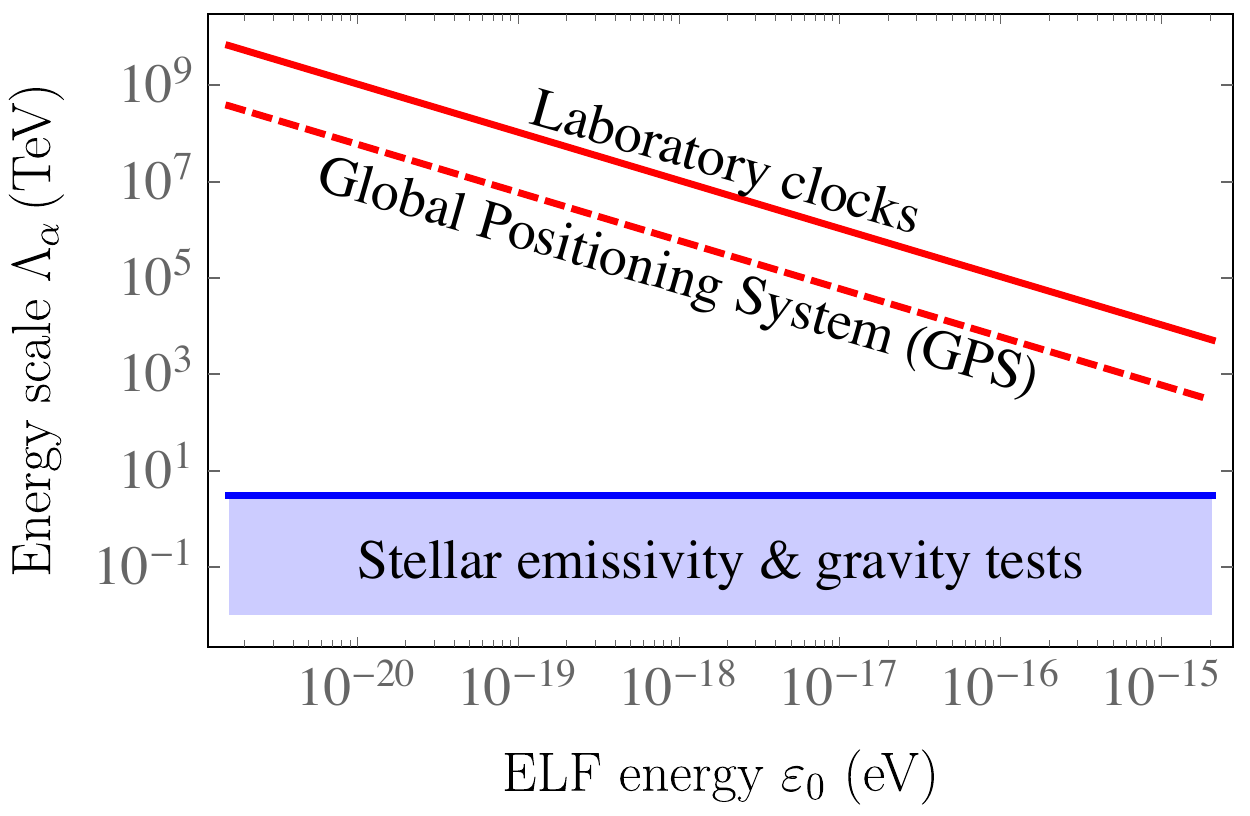}
\caption[]{Projected atomic clock sensitivity to ELFs plausibly emitted during the BNS merger GW170817.  
The discovery reach is shown for a trans-European network of  laboratory clocks (red line, $\sigma_y(1\, \mathrm{s}) =10^{-16}$)  and for the GPS constellation (red dashed line, $\sigma_y(1\, \mathrm{s}) =10^{-13}$).
We assumed an ELF burst of duration $\tau = 100 \,\mathrm{s}$, energy release $0.1 \Msun c^2$, and a total observation time of one month.  Prior constraints~\cite{Olive:2007aj} on the energy scale $\Lambda_\alpha$ are shown by the blue shaded region. From our paper~\cite{dailey2020ELF.Concept}.
 }
\label{Fig:ClockLimits-quad}
\end{figure}

As an example, in Fig.~\ref{Fig:ClockLimits-quad}, we plot the projected sensitivity to a putative ELF burst emitted during the BNS  merger GW170817 ($R=40 \,\mr{Mpc}$). It is clear that existing clock networks can be sensitive to ELFs for a typical GW event (either BNS, BBH or BH+NS mergers) registered by GW detectors. If the sought ELF signal is not observed,
the sensors can place constraints on theoretical models. 

In our GPS.ELF group, we  started data analysis from an atomic clocks  onboard the Global Positioning system (GPS) satellites. The GPS constellation is nominally comprised of 32 satellites in medium-Earth orbit. The satellites house  microwave atomic clocks on-board and they have been used for dark matter searches~\cite{roberts2017search,roberts2018search}.  A network of specialized Earth-based GPS receivers measures the carrier phase of the microwave signals resulting in  the GPS clock time-series data.  One can think of the GPS constellation as the largest human-built $\sim 50,000 \,\mr{km}$-aperture sensor array.  Combined with other satellite positioning constellations and terrestrial clocks, $N_s \sim 100$.  The case of GPS is particularly intriguing as  $\sim 20$ years worth of archival GPS data is available and the dataset is routinely updated~\cite{MurphyJPL2015}.  If an ELF signal is discovered in  recent data, one can go back to pre-LIGO era  and search for similar signals in the archival data. Another  possibility is to correlate the catalogued short gamma ray bursts~\cite{Paul2017} or other powerful astrophysical events with the archival GPS data to search for  ELF bursts. 

Employing networks is crucial for distinguishing ELF signals from technical noise. Furthermore, by having baselines with the diameter of the Earth or larger, one can resolve the sky position of the ELF source. This is a critical feature for multi-messenger astronomy that enables correlation with other observations of the progenitor.   Atomic clocks  have a relatively low $\sim 1 \, \mathrm{Hz}$ sampling rate.  The ELF propagation time across the GPS constellation is $0.2 \, \mr{s}$. The publicly available GPS data is sampled every 
$\Delta_t = 30\, \mr{seconds}$. Our GPS.ELF group has developed techniques to generate higher sampling rate  $\Delta_t = 1\, \mr{second}$ GPS datastreams.
Still the limited time resolution makes tracking the leading edge of the ELF pulse across the GPS network challenging. 
Nevertheless, clock networks can still act collectively, gaining $\sqrt{N_s}$ in sensitivity and vetoing signals that do not affect all the sensors in the network. To mitigate the low sampling rate, one can envision increasing the baseline, similar to recently proposed~\cite{Tino-SAGE-2019} space-based GW detectors. Another possibility is a small-scale ($\sim 10 \, \mathrm{km}$) terrestrial network of optical cavities which allow for $ \gtrsim 10\,  \mr{kHz}$ sampling rate.  The ELF sensitivity of a cavity network is similar to that of the clock networks shown in  Fig.~\ref{Fig:ClockLimits-quad}. 

In summary, we proposed a novel, exotic physics, modality in multi-messenger astronomy.   ELFs serve as messengers and we proposed to employ global networks of precision quantum sensors for their direct detection. This approach must be contrasted with the {\em indirect} exotic-physics detection strategies, e.g., based on minute exotic-physics induced changes in gravitational wave spectral features.  Our strategy benefits from (i) the exquisite sensitivity of atomic quantum sensors and from  (ii)  the enormous amounts of energy released in the mergers. Bursts of exotic fields may, for example, be produced 
during the coalescence of  black hole  singularities. An observation of our predicted ELF signal may yield an experimental signature of quantum gravity unchained in the merger.

\section*{Acknowledgments}
I would like to thank my GPS.ELF and GNOME collaborators for stimulating discussions.
This work was supported in part by the Heising-Simons Foundation and by the U.S. National Science Foundation 
 grants PHY-1912465 and PHY-2207546.

\section*{References}

\begin{thebibliography}{10}
\expandafter\ifx\csname url\endcsname\relax
  \def\url#1{\texttt{#1}}\fi
\expandafter\ifx\csname urlprefix\endcsname\relax\def\urlprefix{URL }\fi
\providecommand{\bibinfo}[2]{#2}
\providecommand{\eprint}[2][]{\url{#2}}

\bibitem{LIGOfirstObservation2016}
\bibinfo{author}{Abbott, B.~P.}\emph{et~al.}
\newblock \bibinfo{title}{{Observation of gravitational waves from a binary
  black hole merger}}.
\newblock \emph{\bibinfo{journal}{Physical Review Letters}}
  \textbf{\bibinfo{volume}{116}}, \bibinfo{pages}{061102}
  (\bibinfo{year}{2016}).

\bibitem{LIGOVirgo-NeutronStarMerger2017}
\bibinfo{author}{Abbott, B.~P.} \emph{et~al.}
\newblock \bibinfo{title}{{GW170817: Observation of gravitational waves from a
  binary neutron star inspiral}}.
\newblock \emph{\bibinfo{journal}{Phys. Rev. Lett.}}
  \textbf{\bibinfo{volume}{119}}, \bibinfo{pages}{161101}
  (\bibinfo{year}{2017}).
\newblock arXiv:\eprint{1710.05836}.

\bibitem{multimessenger2017}
\bibinfo{author}{Abbott, B.~P.} \emph{et~al.}
\newblock \bibinfo{title}{{Multi-messenger observations of a binary neutron
  star merger}}.
\newblock \emph{\bibinfo{journal}{The Astrophysical Journal}}
  \textbf{\bibinfo{volume}{848}}, \bibinfo{pages}{L12} (\bibinfo{year}{2017}).

\bibitem{Coulter2017}
\bibinfo{author}{Coulter, D.~A.} \emph{et~al.}
\newblock \bibinfo{title}{{Swope supernova survey 2017a (SSS17a), the optical
  counterpart to a gravitational wave source}}.
\newblock \emph{\bibinfo{journal}{Science}} \textbf{\bibinfo{volume}{358}},
  \bibinfo{pages}{1556--1558} (\bibinfo{year}{2017}).
\newblock arXiv:\eprint{1710.05452}.

\bibitem{Hallinan2017}
\bibinfo{author}{Hallinan, G.} \emph{et~al.}
\newblock \bibinfo{title}{{A radio counterpart to a neutron star merger}}.
\newblock \emph{\bibinfo{journal}{Science}} \textbf{\bibinfo{volume}{358}},
  \bibinfo{pages}{1579--1583} (\bibinfo{year}{2017}).
\newblock arXiv:\eprint{1710.05435}.

\bibitem{Soares-Santos2017}
\bibinfo{author}{Soares-Santos, M.} \emph{et~al.}
\newblock \bibinfo{title}{{The electromagnetic counterpart of the binary
  neutron star merger LIGO/Virgo GW170817. I. Discovery of the optical
  counterpart using the dark energy camera}}.
\newblock \emph{\bibinfo{journal}{The Astrophysical Journal}}
  \textbf{\bibinfo{volume}{848}}, \bibinfo{pages}{L16} (\bibinfo{year}{2017}).
\newblock arXiv:\eprint{1710.05459}.

\bibitem{Troja2017}
\bibinfo{author}{Troja, E.} \emph{et~al.}
\newblock \bibinfo{title}{{The X-ray counterpart to the gravitational-wave
  event GW170817}}.
\newblock \emph{\bibinfo{journal}{Nature}} \textbf{\bibinfo{volume}{551}},
  \bibinfo{pages}{71--74} (\bibinfo{year}{2017}).
\newblock arXiv:\eprint{1710.05433}.

\bibitem{Margutti2018}
\bibinfo{author}{Margutti, R.} \emph{et~al.}
\newblock \bibinfo{title}{{The binary neutron star event LIGO/VIRGO GW170817 a
  hundred and sixty days after merger: synchrotron emission across the
  electromagnetic spectrum}}.
\newblock \emph{\bibinfo{journal}{The Astrophysical Journal Letters}}
  \textbf{\bibinfo{volume}{856}}, \bibinfo{pages}{L18} (\bibinfo{year}{2018}).
\newblock arXiv:\eprint{1801.03531}.

\bibitem{Mooley2018}
\bibinfo{author}{Mooley, K.~P.} \emph{et~al.}
\newblock \bibinfo{title}{{A mildly relativistic wide-angle outflow in the
  neutron-star merger event GW170817}}.
\newblock \emph{\bibinfo{journal}{Nature}} \textbf{\bibinfo{volume}{554}},
  \bibinfo{pages}{207--210} (\bibinfo{year}{2017}).
\newblock arXiv:\eprint{1711.11573}.

\bibitem{Ruan2018}
\bibinfo{author}{Ruan, J.~J.}, \bibinfo{author}{Nynka, M.},
  \bibinfo{author}{Haggard, D.}, \bibinfo{author}{Kalogera, V.} \&
  \bibinfo{author}{Evans, P.}
\newblock \bibinfo{title}{{Brightening X-Ray emission from GW170817/GRB
  170817A: further evidence for an outflow}}.
\newblock \emph{\bibinfo{journal}{The Astrophysical Journal Letters}}
  \textbf{\bibinfo{volume}{853}}, \bibinfo{pages}{L4} (\bibinfo{year}{2018}).

\bibitem{dailey2020ELF.Concept}
\bibinfo{author}{Dailey, C.} \emph{et~al.}
\newblock \bibinfo{title}{{Quantum sensor networks as exotic field telescopes
  for multi-messenger astronomy}}.
\newblock \emph{\bibinfo{journal}{Nature Astronomy}}
  \textbf{\bibinfo{volume}{5}}, \bibinfo{pages}{150--158}
  (\bibinfo{year}{2021}).
\newblock arXiv:\eprint{2002.04352}.

\bibitem{bini2017deviation}
\bibinfo{author}{Bini, D.}, \bibinfo{author}{Geralico, A.} \&
  \bibinfo{author}{Ortolan, A.}
\newblock \bibinfo{title}{Deviation and precession effects in the field of a
  weak gravitational wave}.
\newblock \emph{\bibinfo{journal}{Phys. Rev. D}} \textbf{\bibinfo{volume}{95}},
  \bibinfo{pages}{104044} (\bibinfo{year}{2017}).

\bibitem{baumann2019probing}
\bibinfo{author}{Baumann, D.}, \bibinfo{author}{Chia, H.~S.} \&
  \bibinfo{author}{Porto, R.~A.}
\newblock \bibinfo{title}{Probing ultralight bosons with binary black holes}.
\newblock \emph{\bibinfo{journal}{Phys. Rev. D}} \textbf{\bibinfo{volume}{99}},
  \bibinfo{pages}{044001} (\bibinfo{year}{2019}).

\bibitem{Loeb2018-BH-signularities}
\bibinfo{author}{Loeb, A.}
\newblock \bibinfo{title}{{Lets talk about black hole singularities}}
  (\bibinfo{year}{2018}).
\newblock arXiv:\eprint{1805.05865}.

\bibitem{JacksonEM}
\bibinfo{author}{Jackson, J.~D.}
\newblock \emph{\bibinfo{title}{{Classical Electrodynamics}}}
  (\bibinfo{publisher}{John Willey \& Sons}, \bibinfo{address}{New York},
  \bibinfo{year}{1999}), \bibinfo{edition}{3rd} edn.

\bibitem{safronova2018search}
\bibinfo{author}{Safronova, M.~S.} \emph{et~al.}
\newblock \bibinfo{title}{Search for new physics with atoms and molecules}.
\newblock \emph{\bibinfo{journal}{Rev. Mod. Phys.}}
  \textbf{\bibinfo{volume}{90}}, \bibinfo{pages}{025008}
  (\bibinfo{year}{2018}).

\bibitem{Olive:2007aj}
\bibinfo{author}{Olive, K.~A.} \& \bibinfo{author}{Pospelov, M.}
\newblock \bibinfo{title}{{Environmental dependence of masses and coupling
  constants}}.
\newblock \emph{\bibinfo{journal}{Phys. Rev. D}} \textbf{\bibinfo{volume}{77}},
  \bibinfo{pages}{043524} (\bibinfo{year}{2008}).
\newblock arXiv:\eprint{0709.3825}.

\bibitem{roberts2017search}
\bibinfo{author}{Roberts, B.~M.} \emph{et~al.}
\newblock \bibinfo{title}{Search for domain wall dark matter with atomic clocks
  on board global positioning system satellites}.
\newblock \emph{\bibinfo{journal}{Nat. Commun.}} \textbf{\bibinfo{volume}{8}},
  \bibinfo{pages}{1195} (\bibinfo{year}{2017}).

\bibitem{roberts2018search}
\bibinfo{author}{Roberts, B.~M.}, \bibinfo{author}{Blewitt, G.},
  \bibinfo{author}{Dailey, C.} \& \bibinfo{author}{Derevianko, A.}
\newblock \bibinfo{title}{Search for transient ultralight dark matter
  signatures with networks of precision measurement devices using a bayesian
  statistics method}.
\newblock \emph{\bibinfo{journal}{Phys. Rev. D}} \textbf{\bibinfo{volume}{97}},
  \bibinfo{pages}{083009} (\bibinfo{year}{2018}).

\bibitem{MurphyJPL2015}
\bibinfo{author}{Jean, Y.} \& \bibinfo{author}{Dach, R.}
\newblock \bibinfo{title}{{International GNSS Service Technical Report 2015
  (IGS Annual Report)}}.
\newblock \bibinfo{type}{Tech. Rep.}
\newblock  (\bibinfo{year}{2016}).

\bibitem{Paul2017}
\bibinfo{author}{Paul, D.}
\newblock \bibinfo{title}{{Binary neutron star merger rate via the luminosity
  function of short gamma-ray bursts}}.
\newblock \emph{\bibinfo{journal}{Mon. Not. R. Astron. Soc.}}
  \textbf{\bibinfo{volume}{477}}, \bibinfo{pages}{4275--4284}
  (\bibinfo{year}{2018}).
\newblock arXiv:\eprint{1710.05620}.


\bibitem{Tino-SAGE-2019}
\bibinfo{author}{Tino, G.~M.} \emph{et~al.}
\newblock \bibinfo{title}{{SAGE: A proposal for a space atomic gravity
  explorer}}.
\newblock \emph{\bibinfo{journal}{Eur. Phys. J. D}}
  \textbf{\bibinfo{volume}{73}}, \bibinfo{pages}{228} (\bibinfo{year}{2019}).
\newblock arXiv:\eprint{1907.03867}.

\end{thebibliography}

\end{document}